\documentclass[twocolumn]{aastex63}

\shorttitle{Mass Transfer in the HD 63021 System}
\shortauthors{Whelan et al}

\begin{document}

\title{HD 63021: Chromospheric Activity and Mass Transfer in a Close
  Binary}

\correspondingauthor{David Whelan}
\email{dwhelan@austincollege.edu}

\author[0000-0002-6006-0537]{David G. Whelan}
\affil{Department of Physics, Austin College, 900 N. Grand Avenue, Sherman, TX 75090}

\author[0000-0003-0346-6722]{S. Drew Chojnowski} 
\affil{Department of Physics, Montana State University, P.O. Box 173840, Bozeman, MT 59717-3840}
\affil{Department of Astronomy, New Mexico State University, PO Box 30001, MSC 4500, Las Cruces, NM 88001}

\author[0000-0002-2919-6786]{Jonathan Labadie-Bartz} 
\affil{Instituto de Astronomia, Geofisica, e Ciencias Atmosfericas, Universidade de Sao Paulo, Brazil}

\author{James Daglen} 
\affil{Daglen Observatory, 33 Joy Rd., Mayhill, NM 88339}

\author{Ken Hudson} 
\affil{Grey Tree Observatory, P.O. Box 290, Mayhill, NM 88339}

\author{Gary M. Casey} 
\affil{Department of Physics, Austin College, 900 N. Grand Avenue, Sherman, TX 75090}
\affil{Department of Applied Physics and Applied Mathematics, Columbia University, 500 W 120th St., New York, NY 10027}

\author[0000-0003-1479-3059]{Guy S. Stringfellow} 
\affil{Center for Astrophysics and Space Astronomy, Department of
  Astrophysical and Planetary Sciences, University of Colorado at
  Boulder, 389 UCB, Boulder, CO 80309-0389}

\author[0000-0002-9903-9911]{Kathryn V. Lester} 
\affil{NASA Ames Research Center, Moffett Field, CA 94035}

\author{John Barry} 
\affil{Department of Physics \& Astronomy, Benedictine College, 1020 N. 2nd St., Atchison, KS 66002}

\author{Joshua Heinerikson} 
\affil{Department of Physics \& Astronomy, Benedictine College, 1020 N. 2nd St., Atchison, KS 66002}

\author{Dan Pankratz} 
\affil{Department of Physics \& Astronomy, Benedictine College, 1020 N. 2nd St., Atchison, KS 66002}

\author{Mary Schreffler} 
\affil{Department of Physics \& Astronomy, Benedictine College, 1020 N. 2nd St., Atchison, KS 66002}

\author{Ryan Maderak} 
\affil{Department of Physics \& Astronomy, Benedictine College, 1020 N. 2nd St., Atchison, KS 66002}

\author{Nick Lotspeich} 
\affil{Department of Mathematics and Physical Sciences, The College of Idaho, 2112 Cleveland Blvd., Caldwell, ID 83605}

\author[0000-0003-1924-7065]{Molly Vitale-Sullivan} 
\affil{Department of Mathematics and Physical Sciences, The College of Idaho, 2112 Cleveland Blvd., Caldwell, ID 83605}

\author{Micah Woodard} 
\affil{Department of Mathematics and Physical Sciences, The College of Idaho, 2112 Cleveland Blvd., Caldwell, ID 83605}

\begin{abstract}

Prompted by X-ray detections from multiple surveys, we investigated
the A-type star HD~63021 and found that it is a double-lined
spectroscopic binary with highly variable emission associated with the
primary star. Analysis of our multi-epoch spectroscopic observations,
the majority of which were carried out on small aperture telescopes,
indicates a very short orbital period of just $2.9$~days, and a mass
ratio M$_2$/M$_1$\ of $0.23$. The A1~V star is a slow rotator, with a
rotational speed of $\sim34$~km/s. Assuming its mass is
$2.3$~M$_{\odot}$, the present-day secondary is an evolved star of
$\sim0.5$~M$_{\odot}$\ that nearly fills its Roche lobe. This
secondary star rotates comparatively rapidly at $\sim44$~km/s, and we
see evidence that it is chromospherically active. Analysis of a
photometric lightcurve from TESS reveals two strong periods, one at
the orbital period for the system and another at half the orbital
period. These findings suggest that HD~63021 is a close binary system
undergoing mass transfer from the secondary star onto the primary star
-- in all ways like an Algol eclipsing binary system, except without
the eclipse. We discuss the system's mass transfer, which is not
steady but seems to run in fits and bursts, and infer the system's
basic physical properties from an orbital parameter study, the Roche
lobe geometry, and its extant X-ray emission.

\end{abstract}

\section{Introduction}\label{intro}

Estimates of the binary fraction of intermediate-mass stars on the
Main Sequence (spectral types B5 down to F2) are typically above
$50$\% \citep[][and references therein; Section 3.4]{Duchene13}; this
fraction is lower for lower-mass stars, and higher for high-mass
stars. Yet in a recent spectroscopic survey, only $15$\% of a target
sample of 20,000 Main Sequence stars showed evidence of binarity, and
most of these were double-lined spectroscopic binaries
\citep[SB2s;][]{ElBadry18}. While many factors contribute to the low
success rate of spectroscopically identifying binary stars, SB2s at
intermediate- and high-mass are particularly difficult to identify
because of the high luminosity of the primary companion, which
obscures spectral features associated with the secondary.

The vast majority of emission-line B-type stars (abbreviated B{\em e}
but including non-supergiant stars of spectral types O9 to A2) are
rapidly rotating and exhibit characteristic emission in Hydrogen
series and low-ionization metal lines that are formed in circumstellar
disks made of gas ejected from the surface of the star. They account
for roughly $15-20$\%\ of all stars within this spectral type range
\citep{Rivinius13}. This percentage is intrinsically variable due to
the transient nature of B{\em e} disks, whereby the emission
signatures may appear or disappear unexpectedly from spectra over
timescales ranging from days to years \citep{Chojnowski17}. Even so,
studies have shown that the binary fraction of B{\em e} stars is the
same as the binary fraction for regular B-type stars
\citep{Oudmaijer10}, so we should expect to find SB2 B{\em e} stars in
the field sample.

Emission-line A-type (A{\em e}) stars are far more rare than B{\em e}
stars and are typically only found among the earlier subtypes A0-A2
\citep[see, e.g.,][]{Jaschek83,Zorec97}. They are also more difficult
to detect because their disks are lower density, as recent
investigations have shown \citep{Vieira17,Shokry18}. A{\em e} and
B{\em e} stars exhibiting X-ray fluxes are attributable to binary
systems with either compact companions
\citep{Podsiadlowski02,Reig11,Wang18} or stellar companions in either
a detached or semidetached state \citep{Singh96}. The X-ray emission
among A-type stars generally is low or undetected \citep{Schroder07},
and for this reason HD~63021 initially caught our attention.

X-ray detections are quite common among emission-line binaries. As
noted above, emission-line sources with compact binary companions
produce X-rays; the source is the circumstellar disk
\citep{Reig11}. Another class of X-ray emitting binaries are those
with chromospherically active late-type companions. RS~CVn stars
consist of at least one late-type component, exhibit Ca~{\sc ii}~H
and~K emission, star spots, ultraviolet and infrared excesses, and are
oftentimes X-ray sources \citep{Singh96}. The (typically) shorter
period early-type Algol eclipsing binaries are also usually X-ray
emitters. X-ray emission in both RS~CVn- and Algol-type binaries comes
from the late-type companions, which are chromospherically active and
rotating quickly \citep[$>5$~km/s;][]{Hall89}.

We first observed HD~63021 in April~2018 as a sample member of
intermediate-mass stars targeted to determine their spectral types. It
was noted, however, that this A1~V star exhibited emission in the
Balmer series and, what is unusual for most B{\em e} stars, the
Fe~{\sc ii}~(42) multiplet. Follow-up spectroscopic observations
revealed that the photospheric absorption lines and Balmer emission
were variable on a period of mere days, and a search through the
available archives additionally revealed that the source had an
appreciable X-ray flux \citep{Whelan18}. Given the likelihood that
HD~63021 would turn out to be a binary, we then set out on a
multi-epoch, high-resolution spectroscopic survey to detect
Doppler-shifted lines and, because of its type, check for Zeeman-split
lines and chemical peculiarities that would suggest the presence of a
magnetic field \citep[e.g.,][]{Chojnowski19}. We also collected
H$\alpha$\ spectra using small telescopes to study the circumstellar
material in detail.

Here we report the findings of this detailed, multi-epoch and
multi-wavelength study of HD~63021. In Section~\ref{data} we describe
the data that we collected on HD~63021, both new observations and from
various archives. In Section~\ref{obsprops} we describe the observed
properties of this binary system, and in Section~\ref{interp} we
interpret our results in light of stars with similar observed
properties. We present our conclusions in Section~\ref{conclusion}.

\section{Observations and Data Reduction}\label{data}

The Observation Log in Table~\ref{obslog} includes information on
spectra secured at four different telescope facilities. Columns
include the observing epoch, observatory code, number of spectra
taken, spectral resolution, and wavelength range. All but one of our
$165$\ individual spectra were secured in the period of one year, from
April of 2018 to March of 2019. (The additional spectrum is a
low-resolution spectrum secured in April of 2020.) For the purposes of
efficiency, we have not listed each of the individual $165$\ spectra
separately in this table, but have instead listed the number of
spectra secured in each calendar month (epoch), where the label
``Apr18'' stands for April of 2018, ``Nov18'' stands for November of
2018, etc. The Observation Log is split into four parts. The first
part lists low-resolution spectra taken at blue-violet wavelengths,
and these observations are described in Section~\ref{dt_AO} below. The
second part lists high-resolution echelle spectra; see
Section~\ref{dt_ARCES}. And the third and fourth parts list
high-resolution long-slit spectra taken with small-aperture telescopes
(Section~\ref{dt_fu}).

This section also includes descriptions of our various photometric
datasets. Section~\ref{dt_TESS} describes the data from the Transiting
Exoplanet Survey Satellite (TESS), and Section~\ref{dt_arch} describes
the various archival datasets that we utilized in our analysis.

\subsection{Low-resolution Spectra}\label{dt_AO}

Low-resolution optical spectra were obtained using an Lhires~III
long-slit spectrograph built by Shelyak, on the Adams Observatory
$24$-inch telescope at Austin College. We use the code ``AO'' to label
observations from the Adams Observatory in Table~\ref{obslog}.

The Lhires~III spectrograph is in the Littrow configuration, so that
there is no magnification of the slit in the focal plane. We utilize a
$35\micron$\ slit to match the seeing conditions at the Adams
Observatory; since the telescope is operating at $f$/8, the slit width
is $1.5\arcsec$\ on the sky. The charge-coupled device (CCD) is housed
in a Finger Lakes Instrumentation Microline camera which can cool to
$60$~K below the ambient temperature. The CCD chip is an e2v~42-10,
with $13.5\micron$\ pixels in an array $512\times2048$. Two different
dispersion gratings were used, with $600$~grooves/mm and
$1200$~grooves/mm, which offer dispersions of $1.1\mathrm{\AA}$/pixel
and $0.54\mathrm{\AA}$/pixel and wavelength ranges of
$2300\mathrm{\AA}$\ and $1125\mathrm{\AA}$\ respectively. The
1200~grooves/mm grating was used most often, and considering the
projected slit width, exhibits a spectral resolution of
$1.4\mathrm{\AA}$, or $\lambda/\Delta \lambda \approx
2700-3500$\ across the spectral range $3820\mathrm{\AA}$ to
$4950\mathrm{\AA}$.

Data reduction and analysis were performed in python using our own
routines. Science images were bias- and dark-subtracted and flatfield
corrected using an integrated flatfield lamp. Spectra were extracted
for a width twice the $\sigma$\ (standard deviation) value determined
by fitting a Gaussian curve to the dispersion profile, and sky
emission was subtracted at each wavelength element based on average
values of the sky on either side of the stellar spectrum. Wavelength
calibration to rest velocity was performed on the science images using
the wavelength positions of known photospheric absorption features,
and was checked using a Neon-Argon discharge tube spectrum and the
positions of two emission lines due to mercury in the sky spectrum
\citep[at $4046.565\mathrm{\AA}$ and $4358.335\mathrm{\AA}$\ in air;
][]{Kramida18}. The signal-to-noise ratio was computed at every
wavelength using the CCD equation \citep{HowellBook}. The final
spectra were divided by the continuum, which was computed by fitting a
cubic spline between hand-chosen continuum points.

\subsection{High-Resolution Spectra}\label{dt_ARCES}

Additional spectra were obtained at high resolution with the
Astrophysical Research Consortium Echelle Spectrograph (ARCES) on the
ARC $3.5$-meter telescope at Apache Point Observatory (label APO in
Table~\ref{obslog}).

The ARCES is permanently mounted at the NA1 port of the 3.5~m
telescope. The ARCES spectra cover the wavelength range
$3,590-10,500\mathrm{\AA}$ at a resolution of $\approx31,500$
($2.5$~pixels), spread over 107 separate orders on a $2048\times2048$
SITe CCD \citep{Wang03}. Data were reduced and extracted using the
IRAF {\it echelle} task, which includes bias subtraction, scattered
light correction, cosmic ray removal, flatfield correction, and
wavelength calibration using the available Thorium-Argon lamp.

\begin{center}
  \begin{deluxetable}{c c c c c}[h!]
    \tablecaption{Optical Spectroscopic Observations\label{obslog}}
    \tablehead{Epoch & Obs.\tablenotemark{a} & No. of  & Resolution      & $\lambda$ Range \\
                     &                       & Spectra &($\mathrm{\AA}$) & ($\mathrm{\AA}$)}
    \startdata
    \multicolumn{5}{c}{Low-Resolution Blue-Violet} \\
    Apr18 & AO & 4  & 1.4 & 3817-4940 \\
    Apr18 & AO & 2  & 2.9 & 3707-6007 \\
    Nov18 & AO & 2  & 1.4 & 3821-4945 \\
    Jan19 & AO & 18 & 1.4 & 3822-4946 \\
    Feb19 & AO & 5  & 1.4 & 3825-4949 \\
    Mar19 & AO & 1  & 1.4 & 3828-4952 \\
    Apr20 & AO & 1  & 1.4 & 3804-4928 \\
    \hline
    \multicolumn{5}{c}{High-Resolution Echelle} \\
    Jan19 & APO & 1 & 0.091 & 3590-10500 \\
    Feb19 & APO & 2 & 0.091 & 3590-10500 \\
    Mar19 & APO & 9 & 0.091 & 3590-10500 \\
    \hline
    \multicolumn{5}{c}{H$\alpha$ Observations} \\
    Apr18 & DO  & 4  & 0.12 & 6487-6645 \\
    Apr18 & GTO & 4  & 0.12 & 6494-6652 \\
    Jan19 & DO  & 10 & 0.12 & 6533-6686 \\
    Feb19 & DO  & 21 & 0.12 & 6532-6687 \\
    Mar19 & DO  & 26 & 0.12 & 6484-6639 \\
    Apr19 & DO  & 18 & 0.12 & 6484-6639 \\
    \hline
    \multicolumn{5}{c}{Si {\sc ii} Observations} \\
    Mar19 & DO  & 37 & 0.12 & 6283-6446\\
    \enddata
    \tablenotetext{a}{Obs.: The observatories where spectroscopy was secured. AO: Adams Observatory; APO: Apache Point Observatory; DO: Daglen Observatory; GTO: Grey Tree Observatory}
  \end{deluxetable}
\end{center}

\subsection{H alpha and Si~{\sc ii} Follow-Up Observations}\label{dt_fu}

Follow-up observations were a collaborative effort between
professional and amateur astronomers, made to study the
H$\alpha$\ emission profile ($\lambda\sim6563~\mathrm{\AA}$) and to
determine the orbital motion of the primary stellar companion using
the Si~{\sc ii} lines at $6347\mathrm{\AA}$\ and
$6371\mathrm{\AA}$. These data were secured at the Daglen and Grey
Tree Observatories (DO and GTO in Table~\ref{obslog}) using Lhires~III
spectrographs on $14$-inch Schmidt telescopes by Celestron with focal
ratio $f$/11. Owned and operated by experienced amateur astronomers,
the telescopes and spectrographs at Daglen and Grey Tree Observatories
are capable of providing research-level science results, as we will
show in Section~\ref{ob_orbparams}.

Data were reduced using the Integrated Spectrographic Innovative
Software (ISIS)\footnote{ISIS is designed specifically for use with
  Shelyak-built spectrographs, and is available at
  \url{http://www.astrosurf.com/buil/ isis-software.html}}, which
includes a standard data reduction procedure, spectral extraction, and
wavelength calibration using the integrated Neon lamp. These follow-up
observations are also logged in Table~\ref{obslog}.

\subsection{TESS Satellite Photometry}\label{dt_TESS}

The Transiting Exoplanet Survey Satellite (TESS) observes a large
portion of the sky for the express purpose of finding exoplanet
transits \citep{Ricker14}. We utilize its multi-epoch photometry to
study the orbital properties of HD~63021. HD~63021 was not
pre-selected for two-minute cadence observations in the first year
(cycle~$1$) of TESS. We extracted the TESS light curve from the
$30$-minute cadence full frame images (FFIs) and performed a principal
component analysis detrending method to remove systematic trends
common to other nearby stars of similar brightness. Comparing the
photometric data of $30$-minute cadence curves extracted this way to
the two-minute light curves delivered by the TESS office for stars of
similar brightness to HD~63021 yields virtually identical results
\citep{LabadieBartz20}. The main differences are that longer term
astrophysical trends remain intact when using the FFIs, and that the
Nyquist frequency for 30-minute cadence data is $24$~c/d, compared to
$360$~c/d for $2$-minute cadence data. There is variability on a scale
many times that of the orbital period that is of uncertain origin; we
present the full light curve in Section~\ref{ob_LC}.

\subsection{Archival Data: Photometric Surveys and GCVS}\label{dt_arch}

Photometry from the Tycho-2 \citep{Hog00}, 2MASS \citep{Skrutskie06},
and WISE \citep{Cutri13} surveys was collected to make a spectral
energy distribution (SED). Flux conversions for the photometry were
performed using the magnitude zero-points found in \citet{Bessell12}
for Tycho-2, \citet{Cohen03} for 2MASS, and \citet{Wright10} for
WISE. Ultraviolet photometry was also collected from the Sky Survey
Telescope \citep[S2/68;][]{Thompson78}. A full table of the observed
fluxes in each band is provided in Table~\ref{sedlog}.

\begin{center}
  \begin{deluxetable}{c c c c c}[h!]
    \tablecaption{Archival Photometric Band Fluxes\label{sedlog}}
    \tablehead{Survey & Band & $\lambda$ & Magnitude & $\nu$F$_{\nu}$ \\
                      &      &($\micron$)&           & (10$^{-10}$ erg/s/cm$^2$)}
    \startdata
    S2/68   & F1565 & 0.1565  & --              & 84$\pm$4\\
    S2/68   & F1965 & 0.1965  & --              & 131$\pm$6\\
    S2/68   & F2365 & 0.2365  & --              & 121$\pm$1\\
    S2/68   & F2740 & 0.2740  & --              & 129$\pm$3\\
    Tycho-2 & B$_T$ & 0.4215  & 7.151$\pm$0.016 & 384$\pm$6\\
    Tycho-2 & V$_T$ & 0.5262  & 7.042$\pm$0.012 & 319$\pm$4\\
    2MASS   & J     & 1.235   & 6.621$\pm$0.020 & 86.5$\pm$1.6\\
    2MASS   & H     & 1.662   & 6.417$\pm$0.042 & 50.9$\pm$2.0\\
    2MASS   & K$_s$ & 2.159   & 6.300$\pm$0.021 & 27.9$\pm$0.5\\
    WISE    & W1    & 3.3526  & 6.205$\pm$0.087 & 9.04$\pm$0.72\\
    WISE    & W2    & 4.6028  & 6.088$\pm$0.031 & 4.08$\pm$0.12\\
    WISE    & W3    & 11.5608 & 6.142$\pm$0.015 & 0.263$\pm$0.004\\
    WISE    & W4    & 22.0883 & 5.933$\pm$0.051 & 0.0476$\pm$0.0022\\
    \enddata
  \end{deluxetable}
\end{center}

HD~63021 has been detected in numerous X-ray surveys, most notably the
ROSAT All-Sky Bright Source Catalogue \citep[RASS/BSC;
][]{Voges99,Haakonsen09}, 2RXS \citep{Boller16}, and the XMM-Newton
Slew Survey \citep{Freund18}. We used the GAIA distance of $167.0 \pm
2.4$pc \citep{BailerJones18}, together with the observed source fluxes
to determine X-ray luminosities in the X-ray bands from the 2RXS and
XMM-Newton surveys. The results are presented in Table~\ref{xraylog}.

\begin{center}
  \begin{deluxetable}{c c c}[h!]
    \tablecaption{X-Ray Flux Measurements for HD~63021\label{xraylog}}
    \tablehead{Survey & Stellar F$_X$  & L$_X$  \\
                      & (10$^{-12}$ erg/s/cm$^2$) & (10$^{27}$ mW) }
    \startdata
    2RXS\tablenotemark{a}       & 2.1$\pm$0.4\tablenotemark{b}  & 7.0$\pm$1.3 \\
    XMM-Newton\tablenotemark{c} & 1.2$\pm$0.6                   & 4.0$\pm$2.0 \\
    \enddata
    \tablenotetext{a}{0.1-2.4~keV}
    \tablenotetext{b}{The flux derived using a power law fit is used, as it had the lowest reduced $\chi ^2$\ value.}
    \tablenotetext{c}{Band 6: 0.2-2~keV}
  \end{deluxetable}
\end{center}

Lastly, the General Catalogue of Variable Stars
\citep[GCVS;][]{Samus17} was cross-matched\footnote{the CDS X-Match
  Service can be accessed at \url{http://cdsxmatch.u-strasbg.fr/}}
with the stellar X-ray catalog of \citet{Salvato18}. This was used to
create a sample of X-ray sources among variable stars with which to
compare HD~63021's X-ray properties.

\section{Data Analysis}\label{obsprops}

\begin{figure*}[ht!]
  \begin{center}
    \includegraphics[scale=0.35]{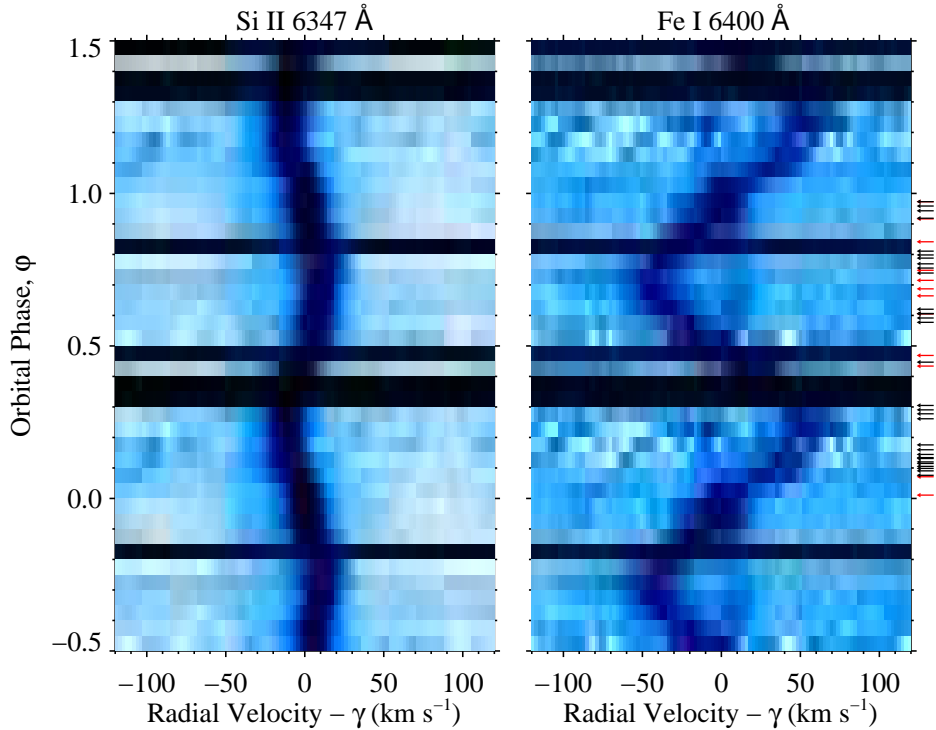}
  \end{center}
  \caption{Orbital phase diagrams for the Si~{\sc ii}
    $6347\mathrm{\AA}$\ line and the Fe~{\sc i}
    $6400\mathrm{\AA}$\ line, where the x-axis is in velocity units,
    and $\gamma$\ is the radial component of the systemic
    velocity. Spectra from the Daglen Observatory were secured at
    phases indicated by the small black arrows to the right of the
    diagram, and spectra from ARCES are indicated by the red
    arrows.\label{fig_SiII}}
\end{figure*}

In this section we analyze our observations and archival data. This
includes a full analysis of our spectroscopic data, starting with
determining orbital parameters from our high-resolution spectroscopic
observations (Section~\ref{ob_orbparams}), identifying emission
signatures of importance (Section~\ref{ob_lineE}), identifying
variable absorption lines in the spectrum, then using absorption lines
to measure projected rotational speed and inferring the equatorial
rotation speed (Section~\ref{ob_absorp}), and finally describing the
H$\alpha$\ line emission variability (Section~\ref{an_Havar}). We then
turn our attention to the photometric datasets, starting with a light
curve analysis (Section~\ref{ob_LC}), then fitting the spectral energy
distribution with stellar photosphere models (Section~\ref{ob_IR}),
and lastly comparing HD~63021's X-ray luminosity and orbital period
with those of other stellar X-ray sources
(Section~\ref{ob_Xray}). Interpretation of these results is reserved
for Section~\ref{interp}.

\subsection{Orbital Parameters}\label{ob_orbparams}

We see direct evidence of this source being a double-lined
spectroscopic binary (SB2) in the ARCES and Daglen Observatory
spectra. Orbital phase diagrams of the Si~{\sc ii} line at
$6347\mathrm{\AA}$ and the Fe~{\sc i} line at $6400\mathrm{\AA}$\ are
shown in Figure~\ref{fig_SiII}. The phases at which each spectrum were
secured are shown with small arrows on the right of the diagrams, with
black arrows designating Daglen Observatory spectra and red arrows
designating ARCES spectra. The color-coding of the actual diagram
traces the brightness of the spectrum, where the contrast between
light and dark for the Si~{\sc ii} line is at the $\sim25$\%\ level,
and for the Fe~{\sc i} line is at the $\sim10$\%\ level. The Si~{\sc
  ii} line, originating with the primary, shows a relatively small
Doppler shift throughout the orbital phase diagram ($\sim10$~km/s),
whereas the Fe~{\sc i} line, from the secondary, shows a much more
substantial Doppler shift ($\sim50$~km/s) and in the opposite
direction to the shift in the Si~{\sc ii} line. In the Fe~{\sc i}
phase diagram, a faint Fe~{\sc i} line from the primary can also be
seen, particularly between phases $0.5-0.8$.

\begin{figure}[h!]
\includegraphics[scale=0.27]{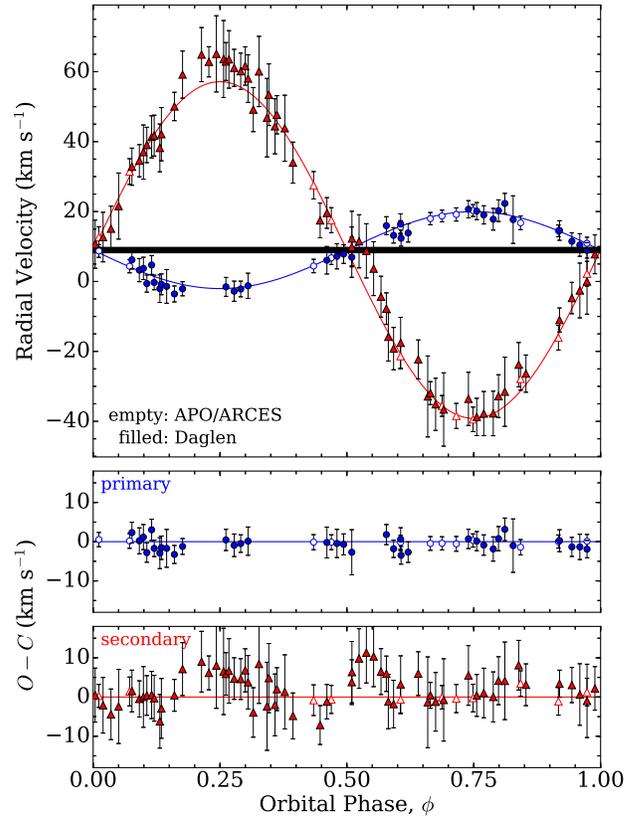}
\caption{Radial velocity for the primary and secondary components as a
  function of phase, with residuals for the primary and secondary
  shown in the lower panels. \label{fig_orb}}
\end{figure}

We identified a number of well-separated lines originating from both
the primary and the secondary stars that could be used to determine an
orbital solution for the system. These include species of O~{\sc i}
($7772\mathrm{\AA}$, $7774\mathrm{\AA}$, and $7775\mathrm{\AA}$),
Mg~{\sc i} ($5167\mathrm{\AA}$, $5173\mathrm{\AA}$, and
$5184\mathrm{\AA}$) and Mg~{\sc ii} ($4481\mathrm{\AA}$), Si~{\sc ii}
($3854\mathrm{\AA}$, $3863\mathrm{\AA}$, and $4130\mathrm{\AA}$),
Ti~{\sc ii} ($4300\mathrm{\AA}$), and Fe~{\sc ii}
($4508\mathrm{\AA}$\ and $4515\mathrm{\AA}$) for the primary, and
Ca~{\sc i} ($6192\mathrm{\AA}$, $6439\mathrm{\AA}$, and
$6463\mathrm{\AA}$) and Fe~{\sc i} ($6180\mathrm{\AA}$,
$6192\mathrm{\AA}$, $6400\mathrm{\AA}$, $6678\mathrm{\AA}$,
$6750\mathrm{\AA}$, and $8824\mathrm{\AA}$) for the secondary. Lines
were fit with Gaussian curves to determine their line centers, and the
average value of the fit line centers for each star were compared to
the rest wavelengths to determine Doppler speeds. The standard
deviation of the mean was also computed, but likely underestimated the
true uncertainties in the average Doppler speeds because we did not
account for the signal-to-noise in the spectra or the airmass at which
the spectra were taken. A qualitative look at the spectra suggested to
us that we adjust the standard deviation values up by $1$~km/s for the
primary and $2$~km/s for the secondary to compensate for these
additional sources of uncertainty; the final uncertainty values are
shown as error bars in Figure~\ref{fig_orb}.  The Doppler speeds with
uncertainties found for each spectrum were then fit using the code
$rvfit$\ \citep{Iglesias15} to determine the orbital parameters of
HD~63021.

\begin{center}
  \begin{deluxetable}{c c}[h!]
    \tablecaption{Fitted and Derived Orbital Parameters\label{orbparamslog}}
    \tablehead{Parameter & Value}
    \startdata
    $P$\ [days]       &  2.906$\pm$0.002     \\
    $T_p$\ [MJD]      &  58540.995$\pm$0.010 \\ 
    $e$               &  0.00 \\
    $\omega$\ (deg)   &  90.0 \\
    $\gamma$\ [km/s]  &  8.979$\pm$0.170 \\
    $K_1$\ [km/s]     &  11.010$\pm$0.509 \\
    $K_2$\ [km/s]     &  48.164$\pm$0.690 \\
    \hline
    $\chi^{2}$            & 50.925 \\
    $rms_{1}$             & 1.561 \\  
    $rms_{2}$             & 4.446 \\  
    $N_{obs}$ (primary)   & 47 \\  
    $N_{obs}$ (secondary) & 73 \\  
    Time span (days)     & 54.759 \\  
    \hline
    $q = M_2/M_1$            & 0.229 $\pm$ 0.011 \\
    $M_1\sin ^3i$ ($M_\odot$) & 0.051 $\pm$ 0.002 \\  
    $M_2\sin ^3i$ ($M_\odot$) & 0.012 $\pm$ 0.002 \\  
    $a_1\sin i$ ($R_\odot$)   & 0.632 $\pm$ 0.029 \\  
    $a_2\sin i$ ($R_\odot$)   & 2.765 $\pm$ 0.040 \\  
    $a  \sin i$ ($R_\odot$)   & 3.397 $\pm$ 0.049 \\  
    \enddata
  \end{deluxetable}
\end{center}

For double-lined spectroscopic binaries, $rvfit$\ solves for the
orbital period ($P$), epoch of periastron passage ($T_p$) which we
present as a modified Julian Date (MJD), eccentricity ($e$), argument
of periastron ($\omega$), radial component of the systemic velocity
($\gamma$), and semi-amplitudes of radial velocity for the primary
($K_1$) and secondary components ($K_2$). Other derived quantities,
such as the mass ratio ($q$) and projected semi-major axis distances
are presented along with the fitted parameters in
Table~\ref{orbparamslog}. Statistical values related to the fit
include the $\chi^2$\ value of the fit, the root-mean-square of the
data minus the model fit for the primary and secondary ($rms_1$\ and
$rms_2$), the number of observations used, and the time span over
which the observations used in the fit were made.

The fit of the radial velocity as a function of phase for a period of
$2.906$~days is shown in the top panel of Figure~\ref{fig_orb}, with
the circle symbols showing radial velocities for the primary component
and the triangle symbols showing the radial velocities for the
secondary component. Open symbols represent data points from ARCES
spectra, whereas filled symbols are from Daglen Observatory
spectra. In the two lower panels, the residuals are plotted separately
for the primary and secondary on the same scale. The residuals for
both components exhibit roughly the same percentage errors.

\begin{figure*}[h!]
\includegraphics[scale=0.32]{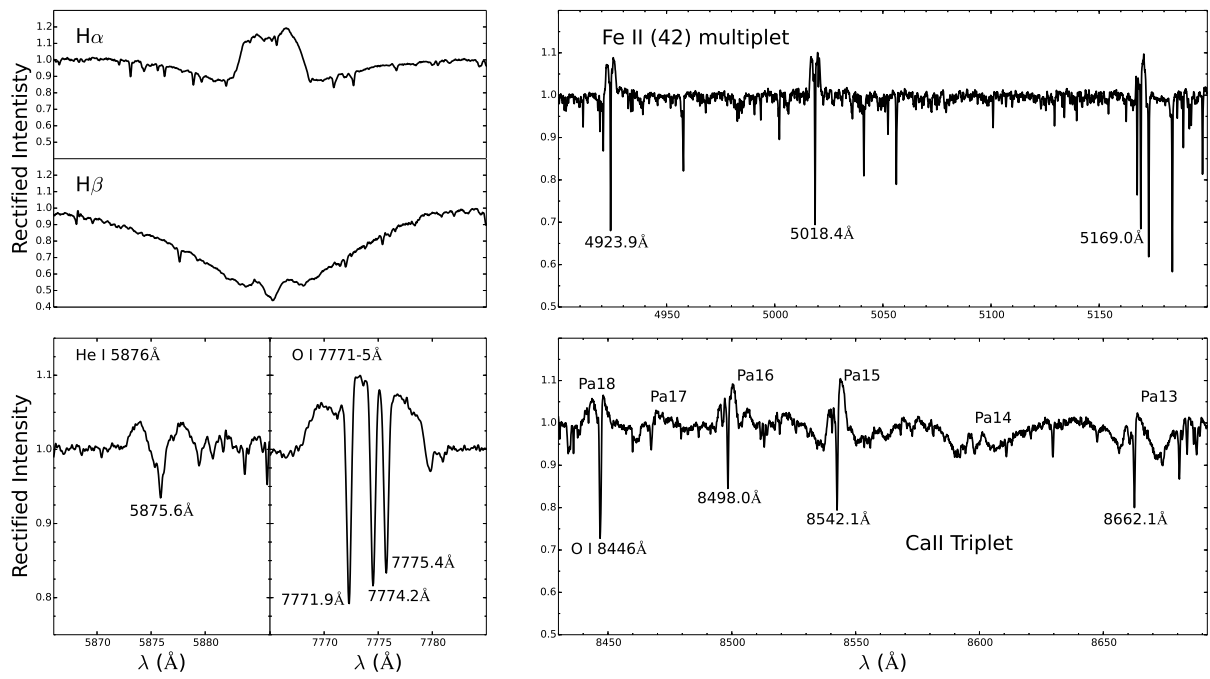}
\caption{Line emission observed with ARCES. Species include H~{\sc i},
  He~{\sc i}, the O~{\sc i} triplet, the Ca~{\sc ii} triplet, and the
  Fe~{\sc ii} (42) multiplet.\label{fig_em}}
\end{figure*}

Given that the spectral type of the primary is A1~V (see below,
Section~\ref{ob_absorp}), if we assume that its mass is
$\sim2.3\pm0.1$~M$_{\odot}$ \citep{Popper80}, then the inclination
angle is $16.3^{\circ} \pm0.3^{\circ}$. This gives us a mass for the
secondary of $0.54\pm0.10$~M$_{\odot}$. We further determine the
semi-major axis $a$\ to be $12.1\pm0.3$~R$_{\odot}$. Using the Roche
lobe effective radius approximation from \citet{Eggleton83}, we
determine Roche lobe sizes of $6.1\pm0.5$~R$_{\odot}$\ for the A1~V
star, and $3.2\pm0.1$~R$_{\odot}$\ for its dimmer and less massive
companion. We use these Roche lobe sizes and the rotational speeds
computed in Section~\ref{ob_absorp} to compute Roche lobe filling
fractions in Section~\ref{interp} below.

\subsection{Atomic Line Emission}\label{ob_lineE}

From the ARCES spectrum taken in January 2019, we show in
Figure~\ref{fig_em} the presence of emission lines from the Balmer
series, He~{\sc i} $5875\mathrm{\AA}$, the O~{\sc i}
$7771-7775\mathrm{\AA}$\ triplet, the Ca~{\sc ii} triplet at
$8498\mathrm{\AA}$, $8542\mathrm{\AA}$, and $8662\mathrm{\AA}$, and
the Fe~{\sc ii} (42) multiplet. The line morphology is similar in
every case: There is a photospheric absorption line that is broad for
the Balmer lines but narrow for the He~{\sc i} and metal lines, with
an emission feature superimposed from the circumstellar material and
exhibiting a double-horned profile as one might expect from a rotating
circumstellar disk. The Fe~{\sc ii}~(42) multiplet and Ca~{\sc ii}
triplet emission is noteworthy also in appearing asymmetric, with the
red emission peak being significantly brighter than the blue. The
O~{\sc i} and Ca~{\sc ii} triplets emission is becoming known as a
tracer of binarity in A{\em e}/B{\em e} stars; this was first studied
in depth in \citet{Polidan76}, and more recently \citet{Chojnowski18}
show them both in the spectrum of the B{\em e}+sdOB binary HD~55606,
and \citet{Banerjee20} confirmed the presence of the Ca~{\sc ii}
triplet in a number of B{\em e} binaries. We also have a growing
dataset of ARCES spectra of B{\em e} binaries that show the same
emission features\footnote{These data are currently unpublished}.

\subsection{The Absorption Line Spectrum}\label{ob_absorp}

\begin{figure*}
\includegraphics[scale=0.3]{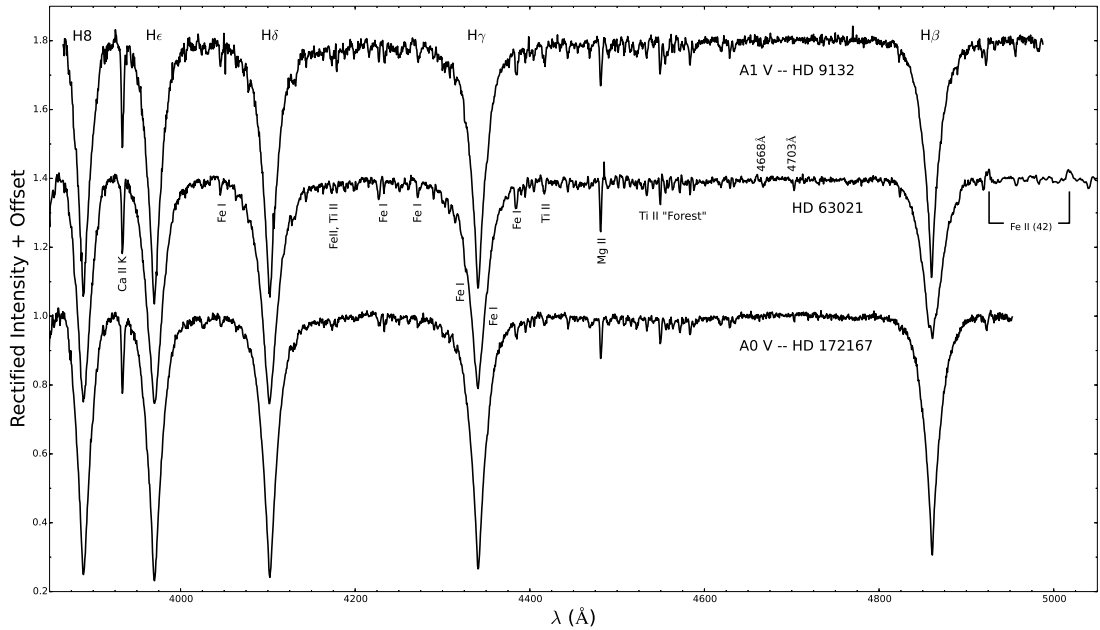}
\caption{Violet-blue spectrum of HD~63021 taken on 2018-04-19 plotted
  against spectroscopic standards of types A0~V and A1~V. Notable
  absorption lines discussed in the text and the Fe~{\sc ii}~(42)
  emission lines are labeled.\label{fig_A1Ve}}
\end{figure*}

A violet-blue spectrum of HD~63021, taken at the Adams Observatory in
April of 2018, is shown in Figure~\ref{fig_A1Ve} along with A0~V and
A1~V spectroscopic standards. The Balmer line wings are closely
matched to the A1~V spectrum, and that is in fact our adopted spectral
type. Filling in of the center of the H$\gamma$\ and H$\beta$\ lines
is clearly visible. The Fe~{\sc ii}~(42) lines indicated at
$4923\mathrm{\AA}$ and $5018\mathrm{\AA}$ are in emission, as
discussed above in Section~\ref{ob_lineE}. The small upward spike near
the Mg~{\sc ii}~$4481$\ absorption line is the result of a hot pixel
that was not corrected during the reduction process. In general the
metallic absorption lines are quite narrow owing to the source's low
$v$sin$i$\ value, and this had to be taken into account when comparing
to the standard spectra.

HD~63021 exhibits some spectral peculiarities. For instance, the
Ca~{\sc ii}~K absorption line in this spectrum more closely matches
the A0~V standard than the A1~V. The various other Fe~{\sc i}, Fe~{\sc
  ii}, and Ti~{\sc ii} lines labeled are by and large at an
intermediate strength between the two standards. Some other lines are
visible at strengths that would normally suggest later spectral types
of A3-5, such as those visible at $4668\mathrm{\AA}$\ and
$4703\mathrm{\AA}$.

Variability observed in the optical spectra is particularly
noteworthy. The emission line strength changed most visibly, and that
altered to what extent the Balmer line centers were filled and how
great the Fe~{\sc ii}~(42) multiplet showed above the continuum. More
surprising, however, was that the metal lines were all observed to
change strength on a spectrum-by-spectrum basis. The Ca~{\sc ii}~K
line, for instance, varied in spectral type from kB9.5 to kA1. Most of
the metal lines varied in strength from A0 to A1. The
$4668\mathrm{\AA}$\ and $4703\mathrm{\AA}$\ lines were usually more
closely matched with an A1 type. It became fairly clear after we
performed the orbital parameters study that the absorption line
variability seen in the low-resolution spectra was somehow tied to the
orbital phase, although the low resolution of these spectra creates
line blending that precluded any definitive analysis. The Balmer line
wings, unlike the metal absorption line spectrum, were not observed to
change spectral type.

We measured $v$sin$i$\ to study the rotational properties of the
primary and secondary stars, using the Fourier transform and
goodness-of-fit produced by the IACOB project \citep{SimonDiaz14}. We
chose well-separated lines in the highest signal-to-noise ARCES
spectrum taken in March of $2019$\ for this purpose. The code
simultaneously measures the projected rotational speed and the
macroturbulent speed. This last parameter is likely a misnomer for hot
stars, as it is not produced by large-scale turbulent motion
\citep{SimonDiaz12}. The primary component of HD~63021 is an A1 star,
and is considered a `hot' star because it is beyond the granulation
boundary \citep[Chp. 17, ][]{Gray1book}. For the primary, it is
therefore not clear what the macroturbulent parameter is fitting,
other than it is not fitting the rotational component to the line's
broadening. Table~\ref{rotlog} summarizes the findings of our
$v$sin$i$\ analysis. The first column lists the absorption lines for
which measurements were made, and the table is split into two sections
to differentiate between absorption lines associated with the primary
and secondary components. The second column, $v$sin$i$ $_{FT}$, is the
calculation of the $v$sin$i$\ by means of a Fourier transform, in
which the macroturbulent velocity has been set to zero
($v_{mac}=0$). The third colum, $v$sin$i$ $_{GOF}$, is the calculation
of the $v$sin$i$\ using the goodness-of-fit method developed by the
authors, for which the macroturbulent velocities listed in column four
of the table ($v_{mac}$) were calculated. In general we note the
closer agreement between the Fourier transform and goodness-of-fit
results for the primary. But for both stars, there is quite a large
spread in values. If we take an average and standard deviation, then
the $v$sin$i$\ for the primary is $9.4\pm1.0$~km/s, and for the
secondary it is $12.4\pm2.0$~km/s.

\begin{center}
  \begin{deluxetable}{c c c c}[h!]
    \tablecaption{$v$sin$i$\ Measurements\label{rotlog}}
    \tablehead{Line ID & $v$sin$i$ $_{FT}$ & $v$sin$i$ $_{GOF}$ & $v_{mac}$ \\
                       &  (km/s)          & (km/s)            & (km/s) }
    \startdata
    \multicolumn{4}{c}{Primary Star} \\
    Si~{\sc ii} 3862 &  7.8 &  8.3 & 12.5 \\
    Fe~{\sc ii} 4508 &  8.3 &  9.0 &  9.9 \\
    Ti~{\sc ii} 4564 &  9.9 &  9.7 &  8.4 \\
    Fe~{\sc ii} 4629 &  7.7 &  9.7 &  9.0 \\
    Mg~{\sc i}  4703 & 10.8 &  9.8 & 10.1 \\
    Fe~{\sc ii} 5276 & 10.6 & 10.0 & 10.0 \\
    Fe~{\sc ii} 5317 &  8.9 & 10.9 & 13.3 \\
    \multicolumn{4}{c}{Average $v$sin$i$: 9.4$\pm$1.0}\\
    \hline
    \multicolumn{4}{c}{Secondary Star} \\
    Ca~{\sc i} 6122 & 14.2 & 15.2 & 16.6 \\
    Ca~{\sc i} 6439 & 12.2 & 14.0 & 17.6 \\
    Fe~{\sc i} 6593 & 12.0 & 14.2 & 16.3 \\
    Fe~{\sc i} 6663 & 13.7 & 13.4 & 15.3 \\
    Fe~{\sc i} 6678 &  9.7 & 11.7 & 12.9 \\
    Fe~{\sc i} 6750 & 11.3 &  7.4 & 21.7 \\
    Fe~{\sc i} 8824 & 12.3 & 12.2 & 13.4 \\
    \multicolumn{4}{c}{Average $v$sin$i$: 12.4$\pm$2.0} \\
    \enddata
  \end{deluxetable}
\end{center}

Given the inclination angle derived from our orbital parameter
solution, this equates to equatorial velocities of $33.5\pm3.6$~km/s
and $44.3\pm7.2$~km/s for the primary and secondary components
respectively. The value for the primary's rotational speed is a little
on the low side compared to other A0/1-type stars \citep{Royer04}, and
in particular it is rotating below the $\sim90$~km/s equatorial speed
above which envelopes are chemically mixed due to meridional
circulation \citep{Charbonneau93}. It is therefore possible that the
metal absorption line strength variability that we observed and
discussed above is at least partly caused by chemical peculiarities in
the photosphere of the primary star. However, the observed line
strength peculiarities are mild at best, and so we do not label the
star as being chemically peculiar.

\subsection{H$\alpha$\ Variability}\label{an_Havar}

\begin{figure*}
\includegraphics[scale=0.32]{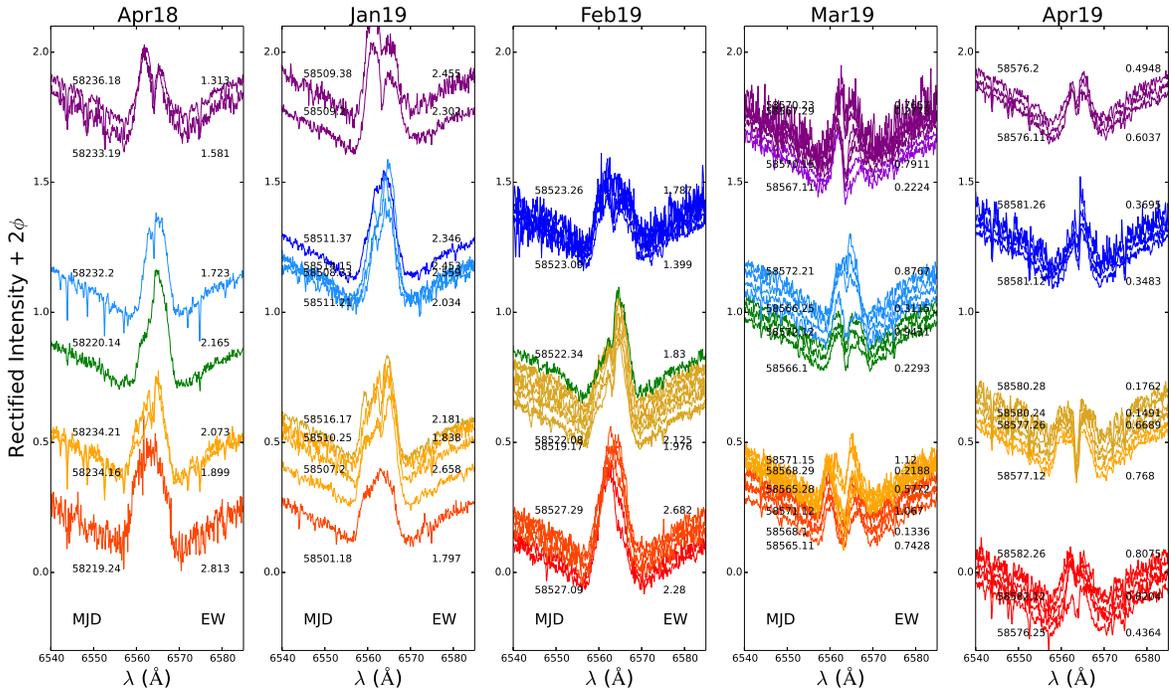}
\caption{H$\alpha$\ observations of HD~63021 over five separate and
  discernible epochs. Each panel shows spectra taken in the epoch
  whose name is shown (Apr18, Jan19, etc. as in
  Table~\ref{obslog}). Spectra are offset according to their phase,
  assuming an orbital period of $2.906$~days (see
  \S~\ref{ob_orbparams}). Modified Julian Dates (MJDs) are given to
  the left of the spectra and emission line equivalent widths (EWs)
  are given to the right; where more than two spectra were obtained in
  a night, only the MJDs and EWs for the first and last observations
  are given for clarity.\label{fig_Ha}}
\end{figure*}

Figure~\ref{fig_Ha} shows our H$\alpha$\ spectra offset by their phase
for each of five months in which it was observed. We plot the modified
Julian date (MJD) on the left-hand side of each spectrum, and the
emission-line equivalent width (EW) measured between
$6556\mathrm{\AA}$\ and $6570\mathrm{\AA}$\ after a local continuum
subtraction on the right. When spectra overlap very tightly, only the
MJDs and EWs for the first and last spectra in a group are shown for
clarity.

In general terms, we note that the H$\alpha$\ emission is highly
variable, but that emission morphologies do repeat roughly on the
orbital period. The emission line strength does change period to
period, and the emission profile morphology can change substantially
after several periods; hence the line morphologies in Feb19 differ
considerably from those in Jan19. During a single period, we see the
line morphology change from asymmetric double-horned profiles, roughly
consistent with a rotating disk plus some extra emission (most
noticeable in the Mar19 and Apr19 epochs) to single-peaked emission,
which is generally when the emission strength is greatest. In what
follows, we present a detailed description of the line profile for
each epoch.

The Apr18 epoch shows rapid variation in emission profile, from a
broad single-peaked line (bottom spectrum), to asymmetric lines with
stronger red components and very small peak separation (middle
spectra), to a stronger blue component and slight wider peak
separation (top spectra). The single-peak profile at the bottom
(MJD$=58219.24$) shows a steeper drop on the redward side than on the
blueward, faintly reminiscent of an inverse P~Cygni profile, which
could suggest infalling material.

\begin{figure}[h!]
\includegraphics[scale=0.5]{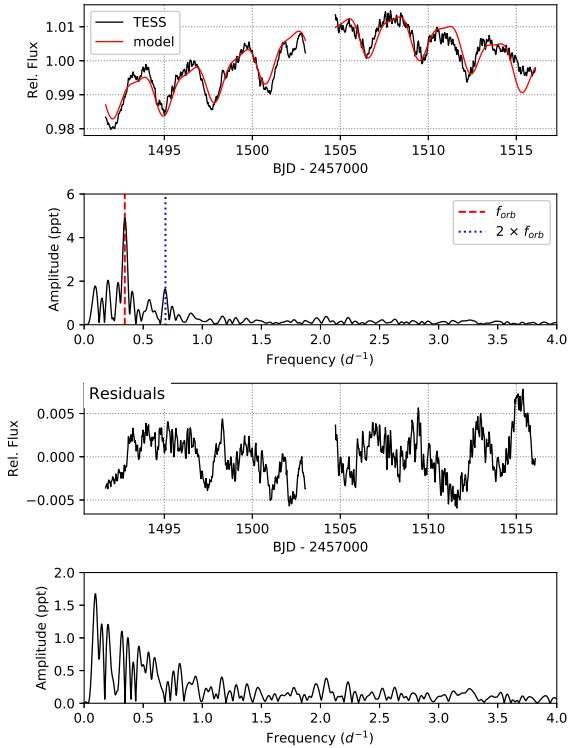}
\caption{The TESS light curve at top is overplotted with a best-fit
  binary model. The periodogram shows that there are strong peaks at
  the orbital frequency and twice the orbital frequency (half the
  orbital period), with the orbital frequency amplitude being much the
  larger. Residuals for the fit are shown, along with a periodogram
  for the residuals that illustrates all orbital motions have been
  accounted for by the model.\label{fig_period}}
\end{figure}

The Jan19 epoch shows roughly similar morphological variations to
those seen in Apr18, with one difference. There is no discernible
pattern in equivalent width measurements between the two epochs: the
single-peak profile at MJD$=58501.18$\ in Jan19 is weaker than the
similar single-peak profile in Apr18, but many (not all) of the
asymmetric profiles that follow have more similar strengths. In short,
line morphology appears decoupled from line strength.

The Feb19 epoch is not as well filled-in as those that precede or
follow it, but the morphological variability is consistent over the
epoch for about eight days, or almost three periods. Line morphologies
look similar to the previous epochs, but shifted in phase by about
-0.25.

Starting in Mar19, the pattern of profile variability changes. The
double-horned profile is a more regular feature, and the peak
separation changes substantially during the orbital period. A similar
pattern of profile variability is evident in the Apr19 epoch, though
the lines are substantially weaker than during any previous epoch and
there is a pronounced central depression between the horns of the
emission profile in about the middle of the period.

\begin{figure}[h!]
\includegraphics[scale=0.4]{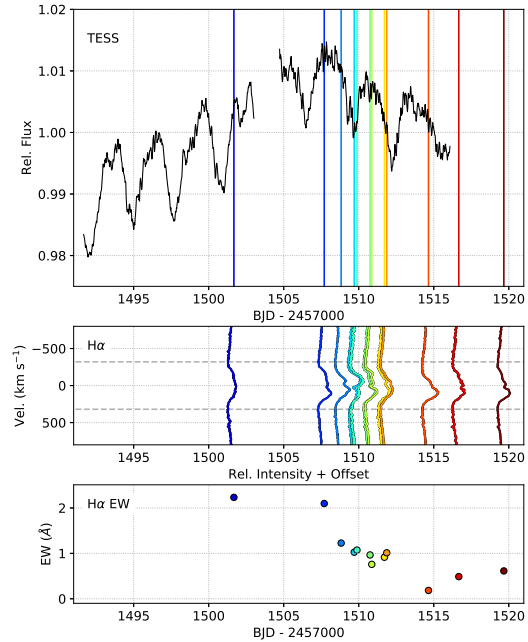}
\caption{{\em Top}: TESS lightcurve, where the colored vertical lines
  represent different ARCES observations. {\em Middle}: The
  H$\alpha$\ line profiles, color-coded to match the lightcurve at
  top. The dotted lines are placed at $6556\mathrm{\AA}$\ and
  $6570\mathrm{\AA}$, inside of which the equivalent widths were
  computed. {\em Bottom}: H$\alpha$\ emission equivalent width
  measurements for each of the ARCES spectra, color-coded with the
  spectra in the middle panel.\label{fig_TESSHa}}
\end{figure}

\subsection{Photometric Lightcurve}\label{ob_LC}

The light curve and periodogram for HD~63021 using TESS data are shown
in Figure~\ref{fig_period}. The light curve is overplotted with a
best-fit model for a binary star system. The periodogram shows clearly
that there are two strong periods associated with the light curve, at
the orbital frequency and at twice the orbital frequency (half the
orbital period). The amplitude of the orbital frequency is almost
three times the amplitude of twice the orbital frequency. Underlying
the brightness fluctuations caused by the orbital motion and
rotational modulation of the system is a gradual brightening and
dimming of the source at a level of $\sim3$\%\ change in
brightness. The residuals for the model-subtracted data show a great
deal of structure at $\sim50$\%\ the level due to the
orbital/rotational signatures, but since no clear period is seen in
the periodogram for these residuals ({\em i.e.}, they appear
stochastic in nature) they are not likely associated with the orbital
motions of the HD~63021 system. The residual frequency spectrum offers
no evidence of further harmonics of the orbital/rotational motion of
the system.

The fact that the strongest amplitude attributable to the orbital
properties of HD~63021 is at the orbital period of the binary is
unusual; generally speaking, the strongest period for binary star
lightcurves is at half the period
\citep{LabadieBartz17,Richards12}. The underlying brightening and then
dimming on a $\sim3$\%\ level for this source requires some
explanation. Its brightening takes place over the course of about
$15$~days, so that if this brightening is indeed periodic it would
take place on a cycle roughly ten times that of the orbital period. It
could be due to brightening of the star itself, or else changes to the
emission associated with the star, since the TESS band includes the
H$\alpha$\ emission line. We plot the TESS lightcurve versus
H$\alpha$\ line profiles observed with ARCES in
Figure~\ref{fig_TESSHa}. Note that the line strength for the
H$\alpha$\ spectrum observed during the source brightening (dark blue
spectrum) is as strong as the emission at peak brightness (second dark
blue spectrum), and that the H$\alpha$\ emission line strength
decreases during the gradual dimming of the source.

\subsection{Spectral Energy Distribution}\label{ob_IR}

\begin{figure}[h!]
\includegraphics[scale=0.4]{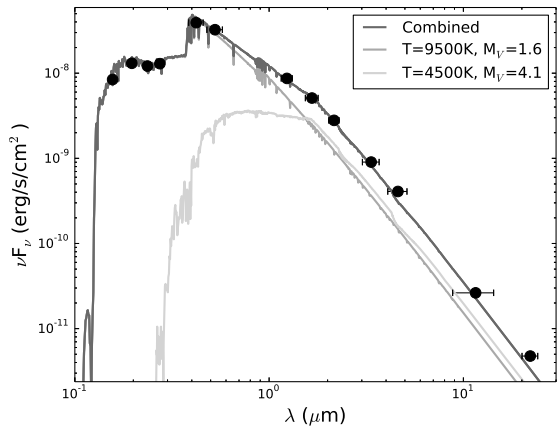}
\caption{Spectral energy distribution (SED) for HD~63021 with model
  spectra co-added and matched in flux to the V-band.\label{fig_sed}}
\end{figure}

We took the collected archival optical and infrared photometry in
order to construct a spectral energy distribution (SED) for HD~63021,
with the express intention of confirming the infrared excess that has
been previously noted by \citet{McDonald12} and \citet{Cotten16}. The
final SED is shown in Figure~\ref{fig_sed}. Model spectra are shown as
solid gray lines for the primary and secondary, and the black line is
the combined spectrum. Photometry from the S2/68, Tycho-2, 2MASS, and
AllWISE catalogues are plotted.

The model spectra are based on the stellar photosphere models of
\citet{Castelli03}\footnote{The model spectra are available at this
  URL: \url{
 https://archive.stsci.edu/hlsps/reference-atlases/cdbs/grid/} \url{ck04models/}}.
We fit the photometry with solar metallicity spectra at T$=9,500$~K
and log$g=4.0$\ for the primary star (A1~V, M$_V=1.6$) and T$=4,500$~K
and log$g=3.5$\ for the secondary star (K5~IV, M$_V=4.1$). The model
choice for the primary comes from matching to the temperature and
gravity recommended by \citet{Graybook} and \citet{Gray1book}
respectively. As for the secondary, there are no model suggestions for
this type. The masses of late-type K dwarfs are close to
$0.5$~M$_{\odot}$, which is why we chose the K5 spectral type. We also
knew it had to be larger than a Main Sequence star because it likely
fills its Roche lobe (see \S~\ref{interp} below); hence, the brighter
magnitude and sub-giant luminosity class.

We assumed intrinsic colors for these two spectral types from
\citet{Fitzgerald70} and absolute magnitudes from
\citet{Graybook}. Using these numbers we determined that no correction
was necessary for interstellar absorption. Given the uncorrected color
of the optical photometry ((B$_T$ - V$_T$)$=0.091$), we did not expect
much interstellar reddening for this source, and our results are
consistent with the {\em maximum} E($B$-$V$) value of 0.013 calculated
by \citet{RuizDern18}. Therefore we did not make a reddening
correction to the photometry.

The UV photometry matches the model photospheric emission very
well. The slight ``knee'' that breaks the Rayleigh-Jeans tail in the
near-infrared data is well-fit by the secondary spectrum. We see a
slight infrared excess above the combined stellar continuum starting
at about $3-4\micron$\ that increases to longer wavelengths,
confirming the infrared excess that has been previously
noted. However, unlike \citet{Cotten16}, who attribute the excess to
circumstellar dust, the excess observed here does not show the typical
strong rise in the mid-infrared that is most often associated with
circumstellar dust emission. Cotten et al. recorded a larger excess
because they fit the photometry with only a single stellar model. The
significantly smaller excess seen here by fitting models of both the
primary and secondary stars is more consistent with free-free emission
by gas, with a power-law slope that is slightly less steep than the
Rayleigh-Jeans tail of the starlight. Excesses similar in shape have
been observed in numerous A{\em e}/B{\em e} systems
\citep[e.g.,][]{Gehrz74}.


\subsection{X-ray Luminosity}\label{ob_Xray}

\begin{figure}[h!]
\includegraphics[scale=0.4]{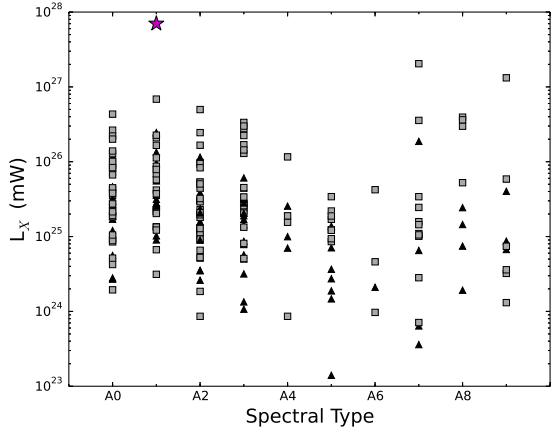}
\caption{X-ray luminosity as a function of spectral type for
  luminosity class IV and V A-type stars from the ROSAT Bright Star
  Catalogue. Known binaries are plotted as squares, while single stars
  are plotted as triangles. HD~63021 is plotted as the large star
  among the A1-type stars. \label{fig_lx_spty}}
\end{figure}

The X-ray luminosity for HD~63021 is high for an A1
star. Figure~\ref{fig_lx_spty} compares the X-ray luminosity for
HD~63021 to luminosities for A-type stars of luminosity classes IV and
V from the \citet{Schroder07} sample. The stars are broken into known
single stars (the black triangles) and known binaries (the gray
squares). On average, binaries have greater X-ray luminosities than do
single stars. Even so, HD~63021 stands out from the crowd.

\begin{figure}[h!]
\includegraphics[scale=0.4]{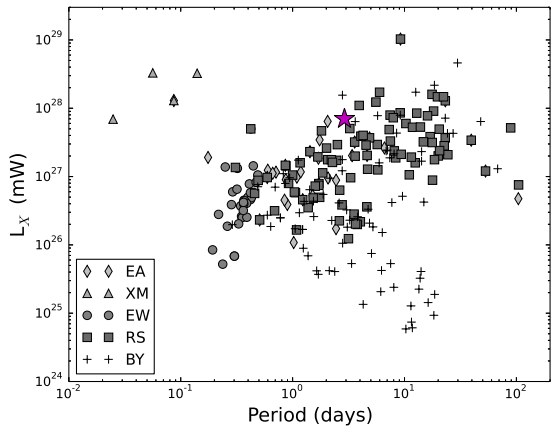}
\caption{X-ray luminosities from the 2RXS survey versus variable star
  period as recorded in the GCVS. See the text for a description of
  the key. HD~63021 is labeled with a star.\label{fig_lx_vars}}
\end{figure}

Since we have computed an orbital period for HD~63021, we next plotted
the X-ray luminosity versus the observed period for variable stars
from the GCVS in Figure~\ref{fig_lx_vars}. This plot shows the X-ray
luminosity and period for some of the most common variability types in
the vicinity of HD~63021: {\bf EA} stands for Algol-type eclipsing
binaries; {\bf XM} stands for variable X-ray emission and the presence
of a strong magnetic field; {\bf EW} stands for W Ursae Majoris
eclipsing binaries; {\bf RS} stands for RS Canum Venaticorum
variables; and {\bf BY} stands for BY Draconis variables. Other
variable stars exist within this parameter space in smaller numbers,
but are not shown. They include $\beta$~Lyrae-type eclipsing binaries,
some types of cataclysmic variables, and some young T~Tauri stars.

Several of these types are unlikely to correspond with HD~63021, based
on the evidence already provided. For instance, HD~63021 is likely not
a W~Ursae Majoris-type binary because those are contact binaries with
much shorter periods and mass ratios closer to unity. It is also not
expected to be a cataclysmic variable because the secondary is not a
compact object, or a T~Tauri star because, given its evolved companion
it is far too old. It is not likely a $\beta$~Lyrae-type binary
because the rapid and steady mass transfer found in those objects
would produce much stronger and steadier Balmer emission. And
BY~Draconis binaries are made up of late-type emission-line stars
only.

The choices that are left in proximity to HD~63021 in
Figure~\ref{fig_lx_vars} are the Algol-type and RS~Canum
Venaticorum-type binaries. Algols are (usually) semi-detached binaries
with early-type (B or early A) Main Sequence primaries and evolved,
lower mass secondaries that are oftentimes losing mass to the
primary. RS~CVn types are also semi-detached binaries, but both
components are usually no earlier than F-type and the evolved
component is the primary. For both Algol and RS~CVn types, the X-ray
emission is due to magnetic activity of the evolved component
\citep{White83}.

\section{Discussion}\label{interp}

HD~63021 is a binary star system exhibiting line emission likely from
a circumstellar disk around one of the components, and X-ray emission
that is consistent with semi-detached systems like Algol-type and
RS~CVn-type binaries. Given the evidence, we wish to determine: (1)
the Roche lobe filling fractions for the primary and secondary stars,
and therefore (2) the location of the emission in the system, as well
as (3) the likely nature of the X-ray emission.

\citet{Abt02} showed that for A-type binaries, the primary star
generally has a rotational speed that is synchronized with the orbit,
so long as the orbital period is less than $4.9$~days. HD~63021's
orbital period is $2.9$~days, so it is reasonable to assume that at
least the primary star is in a synchronized orbit. We now use that
information to determine the size of the A1~V star in HD~63021.

Given the primary's rotation speed of $v_1=33.5\pm3.6$~km/s and
assuming a synchronized orbit, we can compute its radius as
$R=Pv_1$\ where $P$\ is the orbital period derived in
Section~\ref{ob_orbparams}. We determine a radius for the primary of
$R_1=1.9\pm0.2$~R$_{\odot}$. This is spot on when we compare to A1~V
star sizes computed from eclipsing binaries, which inhabit the range
$\sim$1.8-2.2~R$_{\odot}$ \citep{Popper80}. We can perform the same
analysis on the secondary, and assuming that it too is fully
synchronized we determine its radius to be
$R_2=2.5\pm0.4$~R$_{\odot}$. This is slightly smaller than the
secondary's computed Roche lobe size of $3.2\pm0.1$~R$_{\odot}$, which
may indicate that the secondary star is rotating at slightly less than
the synchronous period. Alternatively, there exists a small systematic
uncertainty in our measurements of $v$sin$i$\ for the secondary
(presented in Section~\ref{ob_absorp}), whose lines were very shallow.

In terms of the Roche lobe filling fractions, it is clear that the
secondary star nearly fills its Roche lobe ($\sim70-90$\%), while the
primary has a filling fraction of $0.31\pm0.04$. This value is not
unusual for close binaries with early-type emission-line primary
stars: such a sample studied in \citet{Harmanec15} is composed of
binaries with Roche lobe filling fractions between $0.06-0.48$.

The geometry inferred here is consistent with an Algol-type binary, in
which the more massive primary is the more compact star and is
receiving mass from the secondary \citep{Richards99}, which in this
case appears to be close to but not filling its Roche lobe. As we have
seen, the observed orbital period and X-ray luminosity of HD~63021 is
also consistent with this interpretation (see Figure~\ref{fig_lx_vars}
above). And now we can also say with confidence that the (oftentimes
complex) double-horned profiles seen in many of the emission lines,
including periodically in the H$\alpha$\ line (see
Figures~\ref{fig_em} and~\ref{fig_Ha}), is due to a circumstellar disk
of material around the primary that has been accreted from the
secondary. The line asymmetries suggest that we are not seeing just a
disk, and may be due to emission from the stream of material that
connects the primary's circumstellar disk to L$_1$\ \citep{Lubow75} or
else shock waves created by the interaction between the stream and
disk \citep[e.g.,][]{Bisikalo98,Richards14}. As for the X-ray
emission, \citet{Hall89} has shown that all late-type evolved stars
that are chromospherically active are also rapid rotators, with
rotation speeds in excess of $5$~km/s. The secondary component of
HD~63021 is no exception: its rotational speed of $\sim44$~km/s is
much faster than most late-type stars. This fast rotation speed means
that the corona is denser and hotter than the Sun's, and therefore is
a source for soft X-ray emission \citep{Pallavicini81}.

Having determined the basic properties of the HD~63021 system, we wish
to discuss a few as-yet unanswered features of our data and suggest
resolutions. These include the H$\alpha$\ emission variations, the
longer-period brightening and dimming of the source, the presence of
the He~{\sc i}~$5876\mathrm{\AA}$\ emission line in the spectrum, and
the large residuals exhibited in the model-subtracted light curve.

\begin{figure}[h]
\includegraphics[scale=0.4]{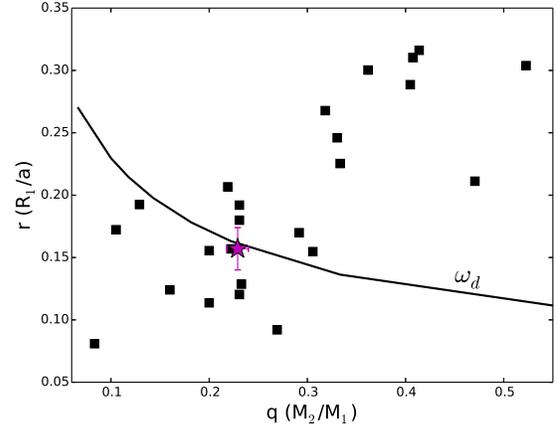}
\caption{The radius of the primary star as a fraction of the star
  separation ($r$) versus the secondary-to-primary mass ratio ($q$)
  for Algol-type binaries. HD~63021 is shown as the
  star. $\omega_d$\ is the smallest distance at which a stable
  circumstellar disk can form as a function of $q$.\label{fig_rq}}
\end{figure}

The H$\alpha$\ emission variations, which are considerable in both
strength and line shape, suggest that the geometry of the
line-emitting regions of the disk and stream are changing over
time. This lead us to wonder whether the accretion disk itself is
dynamically unstable, as is sometimes the case for Algol binaries
\citep{Richards14}. To answer that question, we have plotted in
Figure~\ref{fig_rq} HD~63021's system properties compared to other
known Algol-type binary star systems. Data are taken from
\citet{Richards99} and \citet{Popper80b}, and the parameters plotted
are the radius of the primary as a fraction of the binary separation
($r = R_{*}/a$) and the mass ratio ($q = M_2/M_1$). Also plotted is
the theoretical minimum size of a stable accretion disk as a function
of mass ratio \citep[$\omega_d$; ][]{Lubow75}. HD~63021 lies on the
$\omega_d$\ line, suggesting that any accretion disk created will be
outside this minimum radius and therefore dynamically stable. This
would seem to suggest that the disk geometry is not dynamically
unstable, and left to its own devices would be found in a steady
state. Instead, the {\em amount} of material available is changing,
{\em i.e.} a changing rate of flow from the secondary is the ultimate
cause of the H$\alpha$\ emission variations.

This raises the question: Why does the flow from the secondary star
vary? The photometric light curve (Figure~\ref{fig_TESSHa}) shows a
$\sim3$\%\ rise and fall in source brightness beneath the $2.9$-day
orbital signature. We have additionally noted that the H$\alpha$\ line
strength is roughly correlated with the brightness of the system
(Section~\ref{ob_LC}). These observed features are similar to those of
AU~Mon, for which \citet{Peters94} suggested that the
$1.2$~M$_{\odot}$\ secondary undergoes pulsations that create a cyclic
variation of the mass transfer. Considering the other observed
similarities between AU~Mon and HD~63021 \citep[period-to-period
  variations of the H$\alpha$\ emission as described in
  Section~\ref{an_Havar} and an underlying brightness change on a
  timescale that is about ten times the orbital period, as our data
  suggests;][]{Desmet10}, a cyclic variation in the mass transfer, by
whatever means, is a plausible answer to the observed
H$\alpha$\ emission variations.

Taking the residuals in the light curve (Figure~\ref{fig_period}) at
face value, it seems that there is also a certain amount of stochastic
photospheric activity on either or both of HD~63021's stars. Since
magnetic fields are likely present within the evolved secondary
component, then flares due to the presence of starspots are entirely
possible \citep[e.g.,][]{Hall92}. These starspots may not be
physically large, however, otherwise the periodogram would exhibit a
stronger peak at half the orbital period since the distended shape of
the secondary in its Roche lobe and its synchronous orbit would likely
create starspots on opposite sides of the star \citep{HallHenry92}.

There is still one other possible source of photmetric variation that
is worth briefly noting for completeness, even though we do not see
evidence for it in our light curve. To date, virtually all of the
oscillating Algol-type eclipsing binaries \citep[or
  oEA;][]{Mkrtichian07} have A-type primaries, placing them on the
lower-mass end of the EA group where HD~63021 lies. The pulsational
modes for oEA stars are typically in the $20-150$~minutes range, and
so would appear as a strong period at relatively high frequency. TESS
would certainly have picked up on such a period if it were at least
$60$~mins long (since the observing cadence is $30$~mins; see
Section~\ref{ob_LC}), but no such frequency was observed down to a
noise level of $\sim5\times10^{-5}$.

We turn now to the most unnerving observational trait of HD~63021: the
presence of He~{\sc i}~$5876\mathrm{\AA}$\ emission in the ARCES
spectra. This discovery was unexpected, because the primary star is
not hot enough to produce He~{\sc i} absorption lines in the
blue-violet portion of the spectrum (See Figure~\ref{fig_A1Ve}). The
shape of the emission line is roughly double-horned but with
considerable structure, which suggests that at least some of its
emission originates in the accretion disk around the primary star. If
the line cannot be produced by heating from the photosphere, then the
only other reasonable supposition is heating by shocks. The gas would
be heated by a shock that the stream produces as it interacts with the
disk. Since the accretion flow is unsteady, then the system lacks a
steady-state configuration for its disk and stream, and therefore
shocks are likely at the interface between the stream and disk, as
described in \citet{Lubow75}.

\vspace{1in}

\section{Conclusions}\label{conclusion}

HD~63021 is a double-lined spectroscopic binary in which the
lower-mass secondary is accreting mass onto the A1~V primary -- best
classified as an Algol-type binary, except that it is not
eclipsing. We find an orbital period of $2.9$~days and a mass ratio
$M_2/M_1$\ of $0.23$, and assuming a primary star mass of
$2.3$~M$_{\odot}$\ we compute a secondary mass of about
$0.5$~M$_{\odot}$. The primary star is very likely in synchronous
rotation with the orbit, and we used that to determine the Roche lobe
filling fraction of the primary ($\sim 0.31$). The secondary star,
being the source of the variable circumstellar disk around the
primary, nearly fills its own Roche lobe, and has found some way
(perhaps by means of magnetic activity) to lose mass to the
primary. This mass leaves the secondary's surface and is funneled to
the primary through L$_1$. The photometric lightcurve reveals two
strong periods, at the orbital period and half that value, typical of
such stars, and the orbital period is much the stronger of the
two. The infrared SED exhibits a slight excess, most likely due to
free-free emission from the circumstellar matter, and the X-ray
luminosity is consistent with other binaries containing fast-rotating
chromospherically active late-type stars.

\acknowledgments 

This work would not have been possible without the Adams Observatory
at Austin College. It is also based on observations obtained with the
Apache Point Observatory $3.5$-meter telescope, which is owned and
operated by the Astrophysical Research Consortium. We have made use of
data from the European Space Agency (ESA) mission {\it Gaia}
(\url{https://www.cosmos.esa.int/gaia}), processed by the {\it Gaia}
Data Processing and Analysis Consortium (DPAC,
\url{https://www.cosmos.esa.int/web/gaia/dpac/consortium}). This
publication makes use of data products from the Wide-field Infrared
Survey Explorer, which is a joint project of the University of
California, Los Angeles, and the Jet Propulsion Laboratory/California
Institute of Technology, funded by the National Aeronautics and Space
Administration.

\facilities{Adams Observatory (0.6-meter), ARC, Daglen Observatory,
  Grey Tree Observatory, TESS, HIPPARCOS, FLWO:2MASS, WISE, Sky Survey
  Telescope (S2/68), ROSAT, XMM, Gaia}

\end{document}